# Anomalous diffusion of volcanic earthquakes

SUMIYOSHI ABE[1,2] and NORIKAZU SUZUKI[3]

[1] *Department of Physical Engineering, Mie University, Mie 514-8507, Japan*

[2] *Institute of Physics, Kazan Federal University, Kazan 420008, Russia*

[3] *College of Science and Technology, Nihon University, Chiba 274-8501, Japan*

**Abstract** — Volcanic seismicity at Mt. Etna is studied. It is found that the associated stochastic process exhibits a subdiffusive phenomenon. The jump probability distribution well obeys an exponential law, whereas the waiting-time distribution follows a power law in a wide range. Although these results would seem to suggest that the phenomenon could be described by temporally-fractional kinetic theory based on the viewpoint of continuous-time random walks, the exponent of the power-law waiting-time distribution actually lies outside of the range allowed in the theory. In addition, there exists the aging phenomenon in the *event-time* averaged mean squared displacement, in contrast to the picture of fractional Brownian motion. Comments are also made on possible relevances of random walks on fractals as well as nonlinear kinetics. Thus, problems of volcanic seismicity are highly challenging for science of complex systems.

Volcanic seismicity is a special type of earthquake clustering that occurs not only during but also prior to eruption, indicating its significance as a precursor to such an event. Geophysical approaches to volcanic seismicity involve high-level complexity regarding stress accumulation on faults: dynamics of dikes (propagating, inflating or being magma-filled), nontrivial geometry of the shape of magma migration (such as branching), transport of groundwater through porous media, etc. (see [1-3], for example). Thus, one inevitably faces with complex dynamics on complex architecture.

However, as we will see, volcanic seismicity itself can actually be characterized by a set of surprisingly simple empirical laws.

In this work, we focus our attention on the statistical-mechanical properties of volcanic seismicity. In particular, we perform an analysis in connection with the diffusion phenomenon of volcanic earthquakes, that is, growth of a region in time where earthquakes occur. We report several remarkable findings including the subdiffusive nature, the power-law waiting-time distribution and the exponential-law jump distribution. We make comments on possible relevances of fractional kinetics, nonlinear kinetics and random walks on fractals as well as fractional Brownian motion.

We have analyzed the data of volcanic seismicity at Mt. Etna during 23:34:14 on 12 July, 2001 and 12:46:15 on 8 January, 2011, which is available at http://www.ct.ingv.it/ufs/analisti/catalogolist.php. The region covered is 37.509° N–37.898° N latitude and 14.704° E–15.298° E longitude. The total number of events contained is 5329.

Firstly, we present fig. 1, which shows how epicenters spread in time. There, one can recognize a clear pattern of diffusion. Let $l$ be the radius of a sphere at time $t$, by which all events occurred during the time interval $[0, t]$ are enclosed. Here, the initial time, 0, is adjusted to be the occurrence time of the first event (i.e., 23:34:14 on 12 July, 2001). The spheres are concentric with the center being the hypocenter of the first event. Therefore, the value of $l$ at $t$ is the largest one among the three-dimensional distances between the first event and all subsequent events until $t$. The diffusion property is expressed by the relation

$$l \sim t^{\mu} . \tag{1}$$

$\mu = 1/2$ describes normal diffusion, whereas $\mu \neq 1/2$ is referred to as anomalous diffusion: subdiffusion (superdiffusion) if $0 < \mu < 1/2$ $(1/2 < \mu)$. In fig. 2, we present the plot of $l$ with respect to $t$. There, one finds that eq. (1) holds well with the following small value of the exponent:

$$\mu \cong 0.2 , \tag{2}$$

which implies that volcanic seismicity (at Mt. Etna) is subdiffusive.

The step-like behavior in fig. 2 means that the growth of $l$ is discontinuous: at each step, $l$ remains constant for a certain duration of time. This is in parallel with the concept of *mean maximal excursion* discussed in [4].

The primary result given in eq. (2) necessarily motivates us to examine if the process of the volcanic earthquakes can be described by a kinetic theory. In this respect, we recall that there are (at least) four major approaches to subdiffusion: fractional kinetics based on continuous-time random walks [5-8], fractional Brownian motion [9,10], nonlinear kinetics [11,12] and random walks on fractal structure [13]. For up-to-date discussions about anomalous diffusion, see [14].

To examine the above result based on physical kinetics, we analyze both spatial and temporal statistical properties of this seismicity as a stochastic process.

In fig. 3, we present the plot of the jump probability distribution, $P_J(\rho)$, where $\rho$ is the three-dimensional distance between two successive events [15]. There, we see that it well obeys

$$P_J(\rho) \sim \exp(-\rho / \rho_0) \tag{3}$$

with $\rho_0 = 8.5$ km. This exponential law implies that the spatial property of the process is rather trivial, that is, long jumps are not significant at all unlike in Lévy flights [16].

On the other hand, the temporal property turns out to be nontrivial. In fig. 4, we present the plot of the waiting-time distribution, $P_W(\tau)$, where $\tau$ is the time interval between two successive events (also called "interoccurrence time" or "calm time") [17,18]. Remarkably, it obeys a power law in a wide range:

$$P_W(\tau) \sim \tau^{-1-\alpha}, \tag{4}$$

apart from the rapid decay in the tail region possibly due to finite data-size effects.

Combination of subdiffusion and a power-law waiting-time distribution might seem to indicate that the phenomenon could be described by fractional kinetics based on continuous-time random walks [5-8]. With the waiting-time distribution of the form in eq. (4), a fractional diffusion equation reads, $\partial p(\mathbf{r},t)/\partial t = {}_0D_t^{1-\alpha}\left[D^*\,\nabla^2 p(\mathbf{r},t)\right]$, in the continuous approximation, where $p(\mathbf{r},t)\,d^3\mathbf{r}$ is the probability of finding the walker in the region $[\mathbf{r},\mathbf{r}+d\mathbf{r}] \equiv [x, x+dx]\times[y, y+dy]\times[z, z+dz]$ at time $t$, $D^*$ is a generalized diffusion constant and ${}_0D_t^{1-\alpha}$ is the Riemann-Liouville fractional differential operator [19] defined by ${}_0D_t^{1-\alpha}\left[f(t)\right] = \left\{1/\Gamma(\alpha)\right\}\partial/\partial t\int_0^t ds\,(t-s)^{\alpha-1} f(s)$, provided that the condition $0<\alpha<1$ has to be satisfied. Since the jump distribution is not of the Lévy type, the Laplacian on the right-hand side does not have to be fractionalized. The diffusion property in this case is $l \sim t^{\alpha/2}$ (see recent papers [20,21] and the references therein). This approach is however invalid in the present case, since as can be seen in fig. 4, $\alpha$ in eq. (4) is negative:

$$\alpha \cong -0.13\,, \tag{5}$$

implying that $\alpha$ is outside of the allowed range, $0<\alpha<1$, in fractional kinetics. In addition, if the power law in eq. (4) is extrapolated to $\tau\to\infty$, then $P_W(\tau)$ becomes even unnormalizable. Therefore, a naive picture of fractional kinetics does not apply, here.

Next, let us examine fractional Brownian motion [9,10]. This approach is succinctly described as follows. Let $B_H(t)$ be a process of fractional Brownian motion. It is given in terms of the ordinary Brownian motion $B(t)$ as follows: $B_H(t) = t_*^{-H+1/2}\;{}_0I_t^{H+1/2}\,[\xi(t)]$, where $t_*$ is a positive constant having the dimension of time, $\xi(t)dt = dB(t)$ with $\xi(t)$ being the unbiased Gaussian white noise and ${}_0I_t^{\alpha}$ is the Riemann-Liouville fractional integral operator [19] defined by ${}_0I_t^{\alpha}[f(t)] = \{1/\Gamma(\alpha)\}\int_0^t ds\,(t-s)^{\alpha-1} f(s)$. $H$ is a parameter called the Hurst exponent and should satisfy $0 < H < 1$. Clearly, $B_{1/2}(t) = B(t)$. (For the sake of simplicity, we are discussing in a single spatial dimension, but generalization to higher dimensions is straightforward.) Fractional Brownian motion has self-similarity, $B_H(\lambda t) \approx \lambda^H B_H(t)$, for $\lambda > 0$, where the symbol "$\approx$" stands for equality in the distributional sense, showing monofractality of the process. The sequence of increments $\{B_H(t+t_0) - B_H(t)\}_t$ with $t_0 > 0$ is stationary and strongly correlated. The square root of the variance of the walker's position, $x_H(t) \equiv B_H(t)$, identified with $l$ in eq. (1) gives rise to $\mu = H$. So, subdiffusion is realized in the range $0 < H < 1/2$. Temporal nonlocality yielding long-term memory may be implemented through the fractional integral, although it is not directly connected to the power-law waiting-time distribution in eq. (4). An important feature of fractional Brownian motion is that it does not exhibit the aging phenomenon [14,22] since the sequence of increments is stationary. This point can be examined with the help of the time-averaged mean-squared displacement of the process of the series of the hypocenters, $\{\mathbf{r}_{a+m}\}_{m=0,1,2,\dots,N}$:

$$\overline{\delta^2(n; a, N)} = \frac{1}{N-n} \sum_{m=a}^{a+N-n} \left(\mathbf{r}_{m+n} - \mathbf{r}_m\right)^2, \qquad (6)$$

where $a$, $n$ and $N$ are *aging time*, *lag time* and *measurement time*, respectively, with $N - n >> 1$ being satisfied. We note that in eq. (6) we are using discrete "event time" [23] (see also [24] in the context of complex networks of earthquakes), instead of conventional continuous time in [14,22] (for stationarity of the sequence of increments in discrete-time fractional Brownian motion, see [25]). In fig. 5, we present the plots of the quantity in eq. (6) for several values of $a$. There, the aging phenomenon is clearly observed, in contrast to fractional Brownian motion.

The aging observed in fig. 5 is similar to those in other approaches to anomalous diffusion [14], including continuous-time random walks although already ruled out due to eq. (5).

Finally, we make some comments on other theoretical approaches. Nonlinear kinetics [11,12] is obviously worth being considered if transport of groundwater through a porous medium is taken into account. However, the standard nonlinear diffusion equation known as the porous-medium equation based on Darcy's empirical law and a polytropic relation is local in time and does not reflect the temporal nonlocality highlighted in the power-law nature of the waiting-time distribution in eq. (4). The picture of random walks on fractals [13] is equally attractive if nontrivial geometry of the shape of magma migration [1] such as branching is also taken into account. Subdiffusion in this case comes from correlation between jumps that is induced by fractal geometry, and the resulting process is non-Markovian. Typically, $\mu$ in eq. (1) is given by $\mu = 1/d_w$, where $d_w$ is the walk dimension larger than two [13]. It is however yet to be clarified how the present discoveries of the power-law waiting-time distribution and the aging phenomenon are in accordance with fractal structure inside the volcano.

In conclusion, we have studied the physical properties of volcanic seismicity at Mt. Etna from the viewpoint of kinetic diffusion. We have discovered that it is subdiffusive. We have also analyzed spatio-temporal statistics of this seismicity as a stochastic process and have found that the waiting-time distribution follows a power law in a wide range, whereas the jump distribution obeys a normal exponential law. However, the exponent of the power-law waiting-time distribution has turned out not to be in the range allowed in the theory of fractional kinetics. We have also discussed the event-time averaged mean-squared displacement of the process and have ascertained that it exhibits the aging phenomenon, in contrast to the picture of fractional Brownian motion. We have also made comments on random walks on fractals as well as nonlinear kinetics. It is however fair to say that these possibilities are yet to be further investigated with suitable generalizations or even combinations of them. There, a key seems to be in clarifying if the process is (non-)Markovian. In the case of aftershock sequences in non-volcanic seismicity, there exists a theoretical method for examining if a process is (non-)Markovian. The method employed there uses the concept of singular Markovian scaling relation [26]. However, such a method is not applicable in the present case, since $\alpha$ in eq. (5) is not in the range premised there. So far, no theories seem to successfully describe the phenomenon. Actually, it is a highly nontrivial issue to identify which theory is the correct one for describing observed subdiffusion, in general [27,28]. The present discoveries show how volcanic seismicity is challenging for science of complex systems.

* * *

The authors would like to thank Filippos Vallianatos for drawing their attention to volcanic seismicity at Mt. Etna. SA was supported in part by a Grant-in-Aid for Scientific Research from the Japan Society for the Promotion of Science and by the Ministry of Education and Science of the Russian Federation (the program of competitive growth of Kazan Federal University). NS acknowledges the support by a Grant-in-Aid for Fundamental Scientific Research from College of Science and Technology, Nihon University.

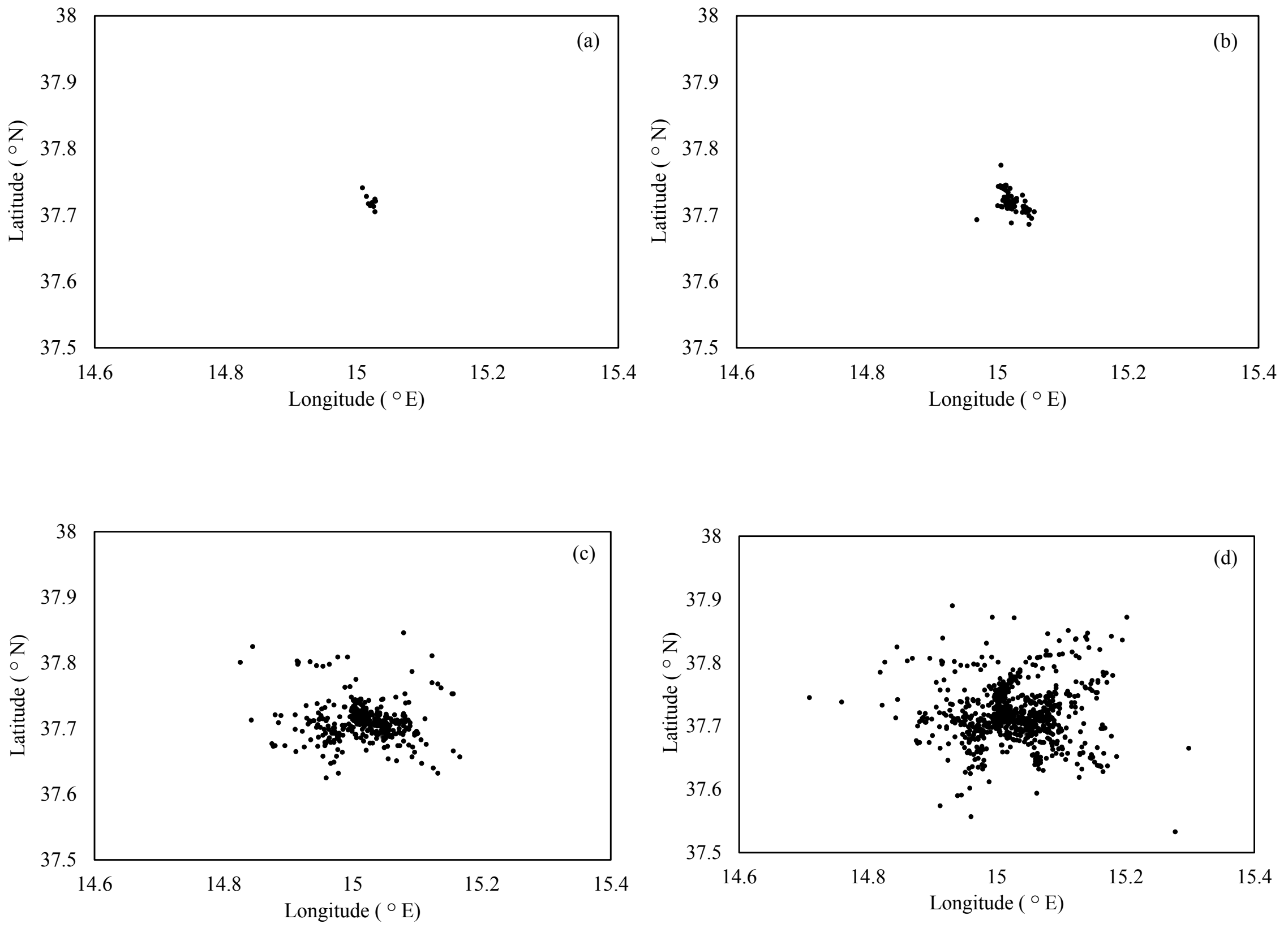

Fig. 1: The epicenters of the first (a) 10 events, (b) 100 events, (c) 500 events and (d) 1000 events contained in the period under consideration.

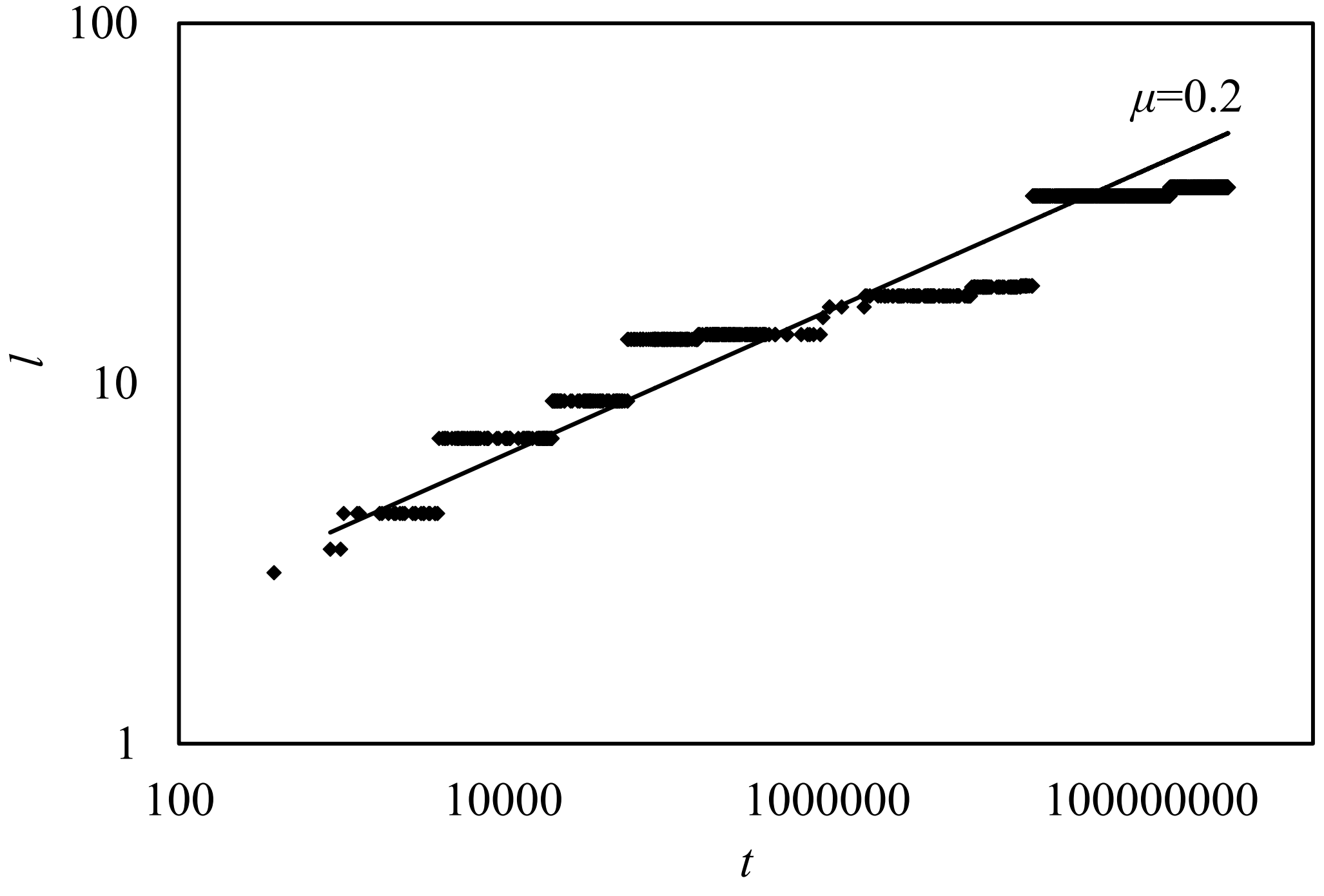

Fig. 2: The log-log plot of $l$ (km) with respect to $t$ (sec).

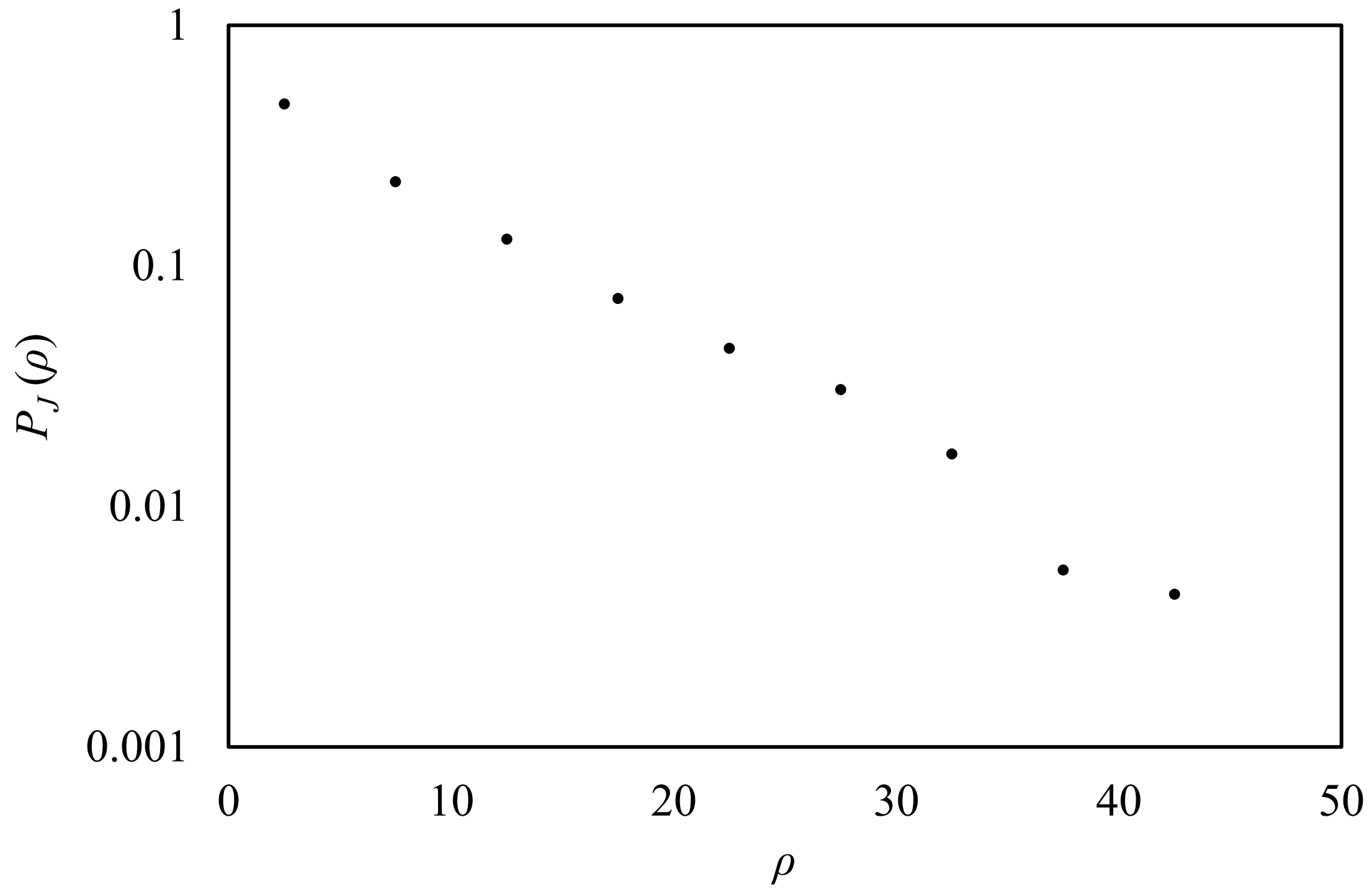

Fig. 3: The semi-log plot of the normalized jump distribution $P_J(\rho)$ with respect to the three-dimensional distance $\rho$ (km) between two successive events. The histogram is made with the bin size of 5 km.

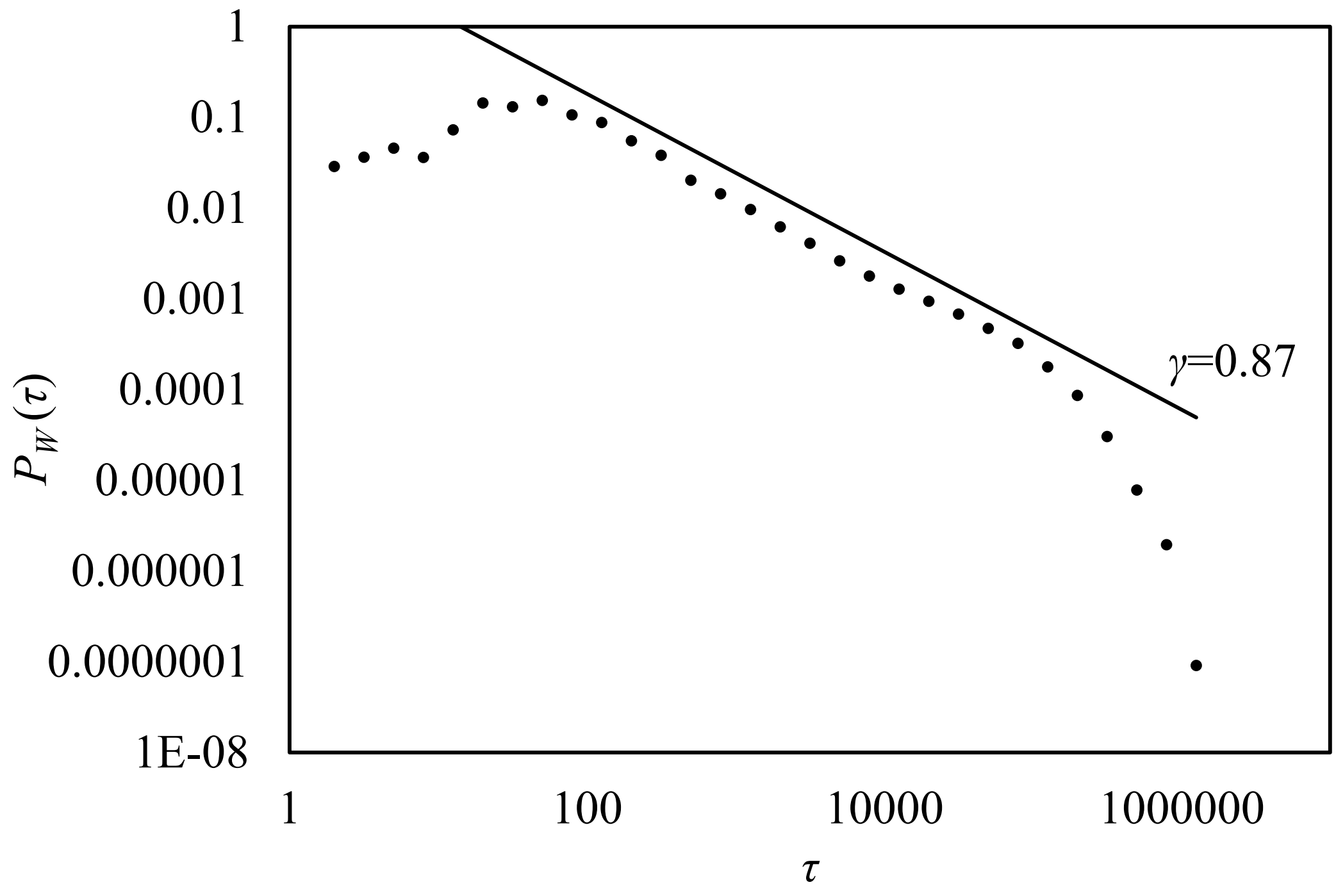

Fig. 4: The log-log plot of the normalized waiting-time distribution $P_W(\tau)$ with respect to the time interval $\tau$ (sec) between successive events with the reference straight line $\tau^{-\gamma}$. The bin size for making the histogram is chosen in such a way that there are five points in each single order of magnitude.

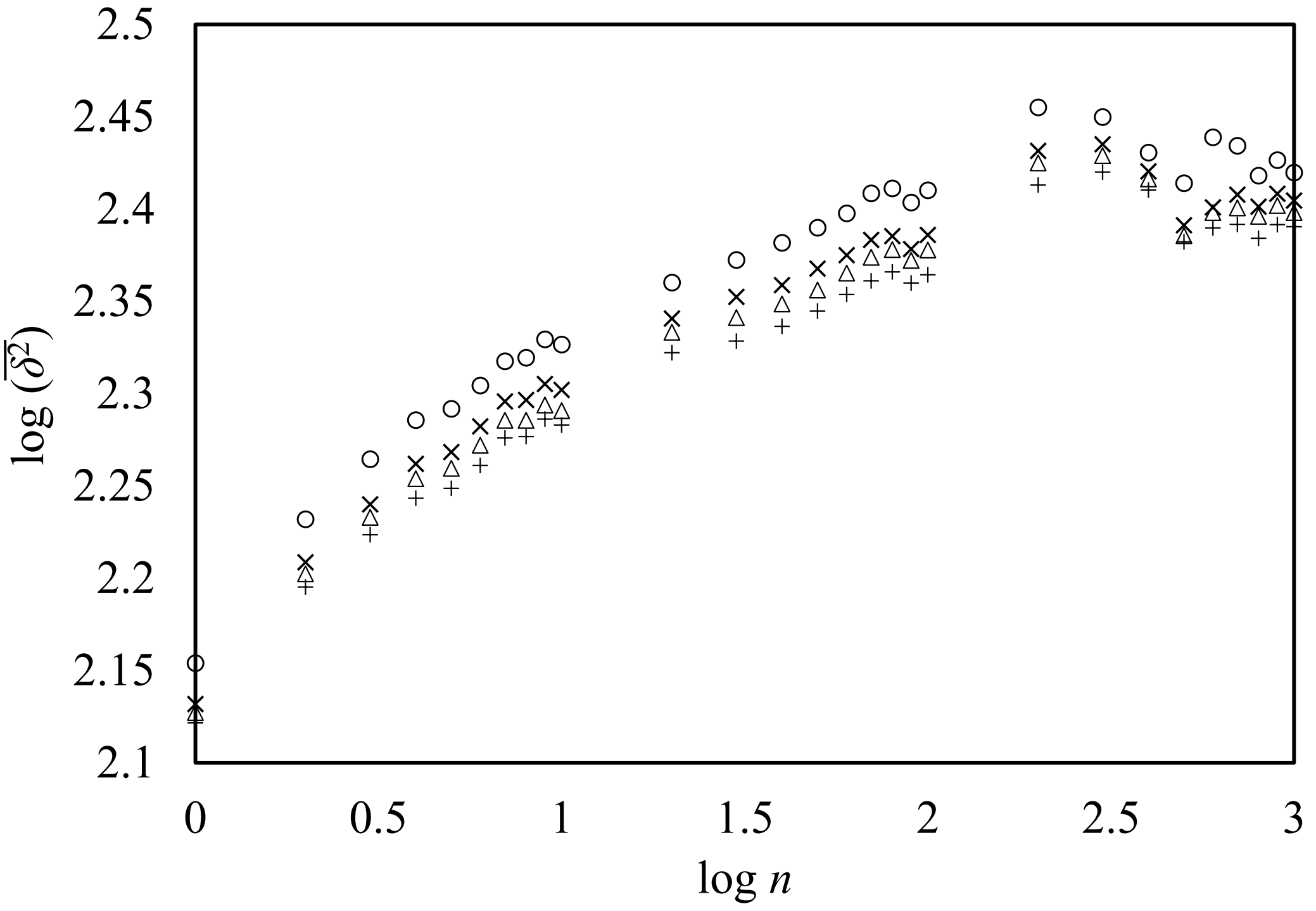

Fig. 5: The plots of the logarithm of $\overline{\delta^2(n; a, N)}$ ($\mathrm{km}^2$) with respect to the logarithm of dimensionless lag event time $n$ for four different values of dimensionless event aging time $a$: + ($a = 0$), $\triangle$ ($a = 50$), $\times$ ($a = 100$), and $\circ$ ($a = 300$). $N$ is taken to be 5000, here.